\documentclass[aps, prl, twocolumn, amsmath, graphicx, latexsym, asmsymb, amsfonts]{revtex4}

\usepackage{epsfig}
\usepackage{graphicx}
\usepackage{epsfig}

\newcommand{\ket}[1]{\ensuremath{|#1\rangle}}

\begin{document}

\title{Frustration, Area Law, and Interference in Quantum Spin Models}

\author{Aditi Sen(De)\(^{\star}\), Ujjwal Sen\(^{\star}\),  Jacek Dziarmaga\(^\natural\), Anna Sanpera\(^{\dagger, \ddagger}\), and
Maciej Lewenstein$^{\star, \dagger}$}

\affiliation{$^\star$ICFO-Institut de Ci\`encies Fot\`oniques,
08860 Castelldefels (Barcelona), Spain\\
\(^\natural\)Institute of Physics and Centre for Complex Systems, Jagiellonian University,  30-059 Krak{\' o}w, Poland\\
\(^\dagger\)ICREA-Instituci{\'o} Catalana de Recerca i Estudis Avan{\c c}ats, Lluis Companys 23, 08010 Barcelona, Spain\\
\(^\ddagger\)Grup de F\'isica Te\`orica, Universitat Aut\`onoma de Barcelona, 08193 Bellaterra, Spain
}

\begin{abstract}
We study frustrated quantum systems from a quantum information perspective. Within this approach,
we find that highly frustrated systems do not follow any general ``area law'' of block entanglement, while weakly 
frustrated ones have area laws similar to those of nonfrustrated systems away from criticality.
To calculate the block entanglement in systems with degenerate ground states, typical in frustrated systems, 
we define a 
``cooling'' procedure of the ground state manifold,  and propose a frustration degree and a method to 
quantify constructive and destructive interference effects of entanglement.
\end{abstract}

\maketitle

The study of entanglement in spin lattice models have provided
a novel insight into the complexity of magnetic ordering and quantum phase transitions
in a large variety of models \cite{xxx}.
In particular, 
in  bipartite partitions  of certain
many-body  systems,
the block entanglement (BE) 
evaluated in 
the (gapped) ground 
state (GS) of the system, is proportional to the
area of the block boundary. More complex relations arise at criticality (non-gapped systems).  These generic relations
between entanglement and area are known under
the name of {\it area laws} \cite{onekey_achhey_eikhaney}.

There are several intriguing 
questions concerning  these relations  \cite{duniyar_golmaal_ek_hou}, with respect to 
systems with unique ground states.
Within this context, one may ask whether the scaling of entanglement
can also help to characterize frustrated many body systems \cite{recent_alet}. 
Typically, frustrated systems exhibit
a large  GS degeneracy, or quasi-degeneracy,  and  a rich phase diagram
\cite{dumdum}.
Frustration in spin  systems arise when the GS spin configuration does not
simultaneously satisfy all the constraints imposed by the Hamiltonian; it may be caused by the lattice geometry
(as in Ising antiferromagnets (AF) on a triangular lattice),
or by the presence of disorder (as in spin glasses).   Recently,  the relation of frustration to high-\(T_c\) superconductivity,
as well as the discovery of ``exotic'' frustrated phases has triggered a renewed interest in the subject
\cite{dumdum,kono-ek-gnaer,Melzi00}.
Ultracold atomic gases offer unprecedented ways to control many-body systems, and provide a perfect 
playground to study disordered and frustrated systems. In particular, this should allow for experimental study of effects discussed in this paper
(cf. \cite{ghyama-paper-sob}, see \cite{amader_review} for a review).

The \emph{main thesis} of this paper is that 
highly frustrated systems do not follow any general area law, while weakly 
frustrated systems have area laws similar to those of nonfrustrated systems (away from criticality).
We introduce three tools to characterize frustrated systems and deal with the above thesis: 
(i) a paradigm of a certain ``cooled state'' 
for finding area laws for systems which have a degenerate ground state manifold, including 
frustrated systems,
(ii) a ``frustration degree'' for quantifying frustration
in a system, 
and  
(iii) a method for quantifying constructive and destructive interference of entanglement. 
Using these tools, 
we prove the above thesis for six prototype frustrated 
spin models, 
each having a different type of frustration. 
More specifically, we show that for the Ising model with long-range 
(LR) AF interactions \emph{(1)}, and the LR AF Heisenberg model \emph{(2)}, 
with frustration degrees close to unity, there is a 
complete departure from the usual area law for nonfrustrated systems. 
However, when we consider systems whose frustration degree is much smaller, 
we find areas laws similar to those 
of non-frustrated systems (away from criticality), i.e. the entropy of a block of characteristic length \(L\),
in a system of \(N\) particles in dimension \(D\), scales as \(L^{D-1}\), when \(L^D \ll N\). 
Among this class, we have analyzed a two-dimensional (2D) $J_1-J_2-J_3$ Heisenberg model with 
resonating valence bond (RVB) states as GSs \emph{(3)}, the 
Shastry-Sutherland model (in 2D) \emph{(4)} \cite{Shastry-Sutherland}, the one-dimensional (1D) 
$J_1-J_2$ Heisenberg model at the Majumdar-Ghosh (MG) 
point \emph{(5)} \cite{Majumdar-ebong-Ghosh}, and finally 
an Ising chain with a single disordered interaction \emph{(6)}.



\emph{Cooling procedure.}
To  treat the problem of degeneracy
of the  GS manifold 
and calculate bipartite entanglement on this manifold, we introduce a ``cooling" method.
Suppose that a system, made up of several subsystems (e.g. spins), is described by the Hamiltonian \({\cal H}\).
We choose some suitable  initial \emph{product} state  \( |\Psi_0 \rangle = \prod_i |\psi_i \rangle\) (where 
\(i\) runs over all subsystems)
and ``cool'' \(|\Psi_0 \rangle\) to below a desired energy level \(\mathbb{E}\),
by projecting
$|\Psi_0\rangle$ onto the subspace that is spanned by all energy eigenstates  of \({\cal H}\) whose energies are below \(\mathbb{E}\).
This "quenching" method is a caricature of evaporative cooling of trapped atomic gases, where atoms having energy above a certain value are
removed from the trap.
So, 
the resulting  state is
\begin{equation}
|\Phi_{\mathbb{E}} \rangle= (1/\sqrt{Z})P[\leq \mathbb{E}] |\Psi_0 \rangle,
\end{equation}
 where
$P[\leq \mathbb{E}] =\sum_{\mathbb{E}_i \leq \mathbb{E}} P_i [\mathbb{E}_i]$, with $ P_i [\mathbb{E}_i]$
being the projector onto the eigenstates of energy \(\mathbb{E}_i\), and
\(Z=
\langle \Psi_0| P[\leq \mathbb{E}]
|\Psi_0 \rangle \).
For 
an \(N\)-spin state
$|\Phi_\mathbb{E}\rangle$, we characterize  the entanglement
between \(k\) spins and the rest of the system by the von Neumann entropy
\(E_{k:N-k}= S(\mbox{tr}_k |\Phi_\mathbb{E}\rangle \langle\Phi_\mathbb{E}|)\) --
the unique asymptotic measure of entanglement for pure states in bipartite splittings \cite{Bennett}.
We study the scaling of BE \(E_{k:N-k}\),
for large \(N\), where the initial state \(|\Psi_0 \rangle\) is chosen to maximize
$ E_{N/2:N/2}$ \cite{footnote}.

\emph{Frustration degree.} Before proceeding further, 
we define a \emph{frustration degree} (\({\cal F}\))
for quantum spin models.
Frustration is a classical concept, and thus connected to the classical configurations 
of the quantum system. Given a Hamiltonian \({\cal H}\), and a corresponding ground state \(|{\cal G}\rangle\),
we replace the one-body, two-body, \(\dots\) terms in \({\cal H}\) by the corresponding Ising ones (i.e. \(\sigma^z_i\), 
\(\sigma^z_i \sigma^z_j\), etc.) to obtain \({\cal H}^{{\cal I}}\),
discarding any constant term.
Here \(\sigma^\alpha\) (\(\alpha = x,y,z\)) are the Pauli operators, and \(\vec{\sigma} = (\sigma_x, \sigma_y, \sigma_z)\). 
We find the terms \({\cal H}_f^k\) (\({\cal H}_{nf}^l\)) of \({\cal H}^{{\cal I}}\) that gives rise to positive
(negative or zero) energies in \(|{\cal G}\rangle\). For a given \(|{\cal G}\rangle\),
\({\cal H}^{{\cal I}} = \sum_k {\cal H}_f^k + \sum_l {\cal H}_{nf}^l\). 
We define the frustration degree
of the system described by \({\cal H}\)  as
\({\cal F} =
\mbox{Av}\frac{\sum_k\langle {\cal G}| {\cal H}_f^k | {\cal G} \rangle }{ \sum_l |\langle {\cal G}| {\cal H}_{nf}^l | {\cal G} \rangle |}\),
where the average (Av) is taken over all ground states of \({\cal H}\). Note that the denominator cannot vanish.
In the case of isotropic Heisenberg (\(\vec{\sigma}_i \cdot \vec{\sigma}_j\)) and XY
(\(\sigma^x_i \sigma^x_j+ \sigma^y_i \sigma^y_j\)) interactions, to find \({\cal F}\), one may  replace \(\vec{\sigma}_i\) and
\((\sigma^x_i, \sigma^y_i)\) by SU(2) and SO(2) symmetric
classical vectors,  of appropriate lengths, respectively.
With these
concepts at hand, we move to study the different cases.

\emph{Case 1: Ising model with AF LR interactions: Ising gas.}
The governing Hamiltonian, for \(2m\) spins, is
\(H^{Ising}_{LR}(\lambda) = (J/2m) (S - 2m\lambda)^2\)
\(\equiv (J/2m)
(2\sum_{i,j=1  (i> j)}^{2m} \sigma_i^z \sigma_j^z - 4m\lambda \sum_{i=1}^{2m}\sigma_i^z + \mbox{const.})\), where
 \(S= \sum_{i} \sigma_i^z\),
\(0 \leq \lambda \leq 1\).
In the  frustrated case (\(J>0\)),
\({\cal F} = (1+2\lambda - \lambda^2 -1/m)/(1+\lambda)^2\). For large systems, with \(\lambda =0\), \({\cal F} \approx 1\).
%
For the initial state 
\(|\Psi_0^I\rangle= \prod_i 
(\alpha |0\rangle + \beta |1\rangle)_i\), with $\alpha \beta \ne 0$
(with \(|0\rangle\) and \(|1\rangle\) being the
 eigenstates of \(\sigma^z\) with eigenvalues $\pm 1$, respectively),
the cooled state (which is independent of \(\alpha, \beta\)) is
%
\(|\Phi_{LR}^{Ising}(\lambda)\rangle = \)
superposition of all states with \(m(1+\lambda)\) \(|0\rangle\)'s
and \(m(1-\lambda)\) \(|1\rangle\)'s, where \(2m\lambda\) is a positive integer \(\leq 2m\).
For arbitrary \(2m\) and magnetization $\lambda$,  and for any bipartite splitting \(k:2m-k\),
the reduced  density matrix can be found analytically: for even \(k < m(1-\lambda)\),
\begin{equation}
\rho_k =
\sum_{i=0}^{k} [k,i] [2m-k, m\mu -i] 
          \left|W_{k-i} \right\rangle \left\langle W_{k-i} \right| / [2m, m \mu] , \nonumber
\end{equation}
where  
$\mu=\lambda+1$, \([a,b] = \frac{a!}{(a-b)! b!}\), and
\(|W_j\rangle \) denotes the normalized equal superposition state  of $j |1\rangle$'s  and rest $|0\rangle$'s,
i.e. the (generalized) ``W'' state.
Any other partition can be found similarly.
For a given \(k\),  \(E_{k:2m-k}\) 
increases (and converges to a constant)
with 
$2m$. For a fixed \(m\),
using Stirling approximation,
\(E_{k:2m-k} \approx
- \sum_{i=0}^k e_i \log_2 e_i\), with \(e_i = \frac{1}{2^k} (1+ \lambda)^r (1-\lambda)^{k-r}
[k,i]\),
for large \(m\) and \(k\), such that \(k \ll m\). In the same limit,
\(E_{k:2m-k} \approx \frac{1}{2} \log_2 \left[(1-\lambda^2)k\right]\). 
The leading term to the scaling is $\frac{1}{2}\log_2 k$, independently of $\lambda$
-- a logarithmic divergence (LD) of BE
for all frustrated AF LR Ising gases.
Since this system is effectively in ``infinite" dimensions, there is a clear departure from any ``area law''
(which could be $k^{1-1/D}$ with $D \to \infty$).
Similar divergence is also obtained even if the quenching is performed over
a few low lying energy manifolds.
In contrast, for the ferromagnetic ($J<0$) nonfrustrated LR Ising model
(say, at \(\lambda =0\)),
the state after cooling is 
\((|0^{\otimes 2 m}\rangle + |1^{\otimes 2 m}\rangle)/\sqrt{2}\),
and has 
BE
\(E_{k:2m-k}\)=1, irrespective of 
\(k\) 
and $2m$.

\emph{Case 2: Heisenberg model with AF LR interactions: Heisenberg gas.} 
The 
Hamiltonian, for \(2m\) spins,
is \(H_{LR}^{{\cal H}}= (J/2m)\sum_{i,j=1}^{2m} \vec{\sigma}_i \cdot \vec{\sigma}_j\) (\(J >0\)).
The ground state manifold can be defined by labeling 
the \(2m\) spins as ``black'' (call them \({\cal B}\)) and the remaining as ``white'' (call them \({\cal W}\)), and 
considering all possible ``coverings'' by singlets (valence bonds) from black to white spins.
Given that a spin-1/2 particle has only two orthogonal states, an upper bound 
for BE in the cooled state, for an initial state
  with two states, is a reasonable approximate upper bound. Cooling the initial state 
\(\left|\Psi^{{\cal H}}_0\right\rangle = \prod_{i \in {\cal B}} \left|\psi^{\cal B} \right\rangle_i \) \(\prod_{i \in {\cal W}} \left|\psi^{\cal W} \right\rangle_i\),
we obtain \(\left|\Phi_{LR}^{{\cal H}}\right\rangle\), which is an RVB formed by an equal superposition of all the singlet coverings. 
Consider a ``system'' \({\cal S}\) of  \(k \ll 2m\) spins (containing, say, 
 \(b\) black spins and \(w\) white ones), and the 
remaining ``environment'' 
\({\cal E}\). 
The orthonormal states of \({\cal S}\) 
in a \({\cal S} : {\cal E}\) Schmidt decomposition 
must be fully symmetric under independent permutations in the subsets of black and white spins.
So the basis of symmetric spin multiplets 
\(\left|b/2, m_b \right\rangle_{{\cal B} \cap {\cal S}} \left|w/2, m_w \right\rangle_{{\cal W} \cap {\cal S}}\) will span them, 
where $m_c=-\frac{c}{2},...,\frac{c}{2}$, \(c=b,w\),
so that
there are $(b+1)(w+1)$ symmetric basis states.
The Schmidt decomposition cannot have more terms than the number of symmetric
basis states,
so that 
\(E_{k:2m-k} \leq \log_2(b+1)(w+1)\).
No matter what is $b$ or $w$, any area law of the form $E\sim k^{1-1/D}$ for any 
$D$ (including $D=\infty$) is out of question. 
In the case when $w=k,b=0$ (or $w=0,b=k$), the cooled state is 
\begin{eqnarray}
\sum_{M=-k/2}^{k/2}
|k/2,M\rangle_{{\cal B} \cap {\cal S}} 
 \sum_{p=-m/2}^{m/2}
(-1)^{(p-m/2)}C^{m,k}_{p,M}/\sqrt{m+1} \nonumber \\
 \times |(m-k)/2,p-M\rangle_{{\cal B} \cap {\cal E}} 
|m/2,-p\rangle_{{\cal W}},
\nonumber
\label{Schmidt1}
\end{eqnarray}
where 
\((C^{m,k}_{p,M})^2 =
[k,M+k/2][m-k,p-M+(m-k)/2]/[m,p+m/2]\) defines the Clebsch-Gordon coefficient 
($C^{m,k}_{p,M}=0$ unless 
$0\leq b\leq a$ in all \([a,b]\) involved),
whence one can show, with some algebra, that
the upper estimate 
is saturated.


\emph{Case 3: 2D $J_1-J_2-J_3$ Heisenberg model with RVB ground states.}
Consider a (2D) square lattice of size \(2m\times 2m\) with periodic
boundary conditions (PBC), each site being occupied by a spin-1/2 particle.
It 
is governed by a
two-body \(J_1-J_2-J_3\) AF Heisenberg Hamiltonian,
with exchange constants $J_1$ for nearest neighbors (NN) and on diagonals in the plaquettes,
$J_2$ for inter-plaquette NNs and diagonals, and \(J_3\) for next-nearest neighbors (NNN) and knight's-move-distance-away spins on
horizontal and vertical ladders formed by the plaquettes
\cite{Brijesh_Kumar_zindabad, Mila06w}.
A plaquette is a square of four neighboring spins.
Realization of such a ``quadrumerized'' lattice is being hotly pursued in current experiments with atoms in optical lattices
(see \cite{paredes+bloch}).
For certain choices of the relative strengths of the couplings, it is reasonable to assume \cite{Brijesh_Kumar_zindabad, Mila06w}
that the GS configuration  is formed by
either two horitzontal singlets \(|HH\rangle\) or  two vertical ones \(|VV\rangle\) in each of the plaquettes, with a
fixed density, \(d=s/m^2\), of  $s$ vertical singlets in the whole state of \(m^2\) plaquettes.
A superposition of such states is an example of an
RVB state, whereas ordered configurations of dimers (singlets) correspond to valence bond solids (Peierls order).
If \(d\) is not too close to \(0\) or \(1\), the dimension
of the GS subspace  grows exponentially
with system size: for \(d=1/2\), the degeneracy scales as \(2^{m^2}\).
Note that in contrast to \emph{Cases 1} and \emph{2},
the model here has rather ``medium''-range
interactions, and the frustration degree will correspondingly be smaller.
For instance, for \(J_2 = J_3 =0\), \({\cal F} =1/2\). 

Consider first the scaling \(E_{p:rest}\), for \(p\) \emph{plaquettes}. For the initial state
%
\(|\Psi_0^H\rangle= \Pi_{p=1}^{m^2}(\alpha |HH\rangle + \beta |VV\rangle)_p$  (unnormalized, and \(\alpha \beta \neq 0\)),
the cooled state (which is independent of \(\alpha\), \(\beta\)) is
\begin{eqnarray}
|\Phi_{RVB}^d\rangle= 
&&\sum_{l=0}^{\min(s,k)}
 \sum_{r=0}^{\min(s-l,m^2-k)}
|l\rangle_{\cal S}~
\sqrt{3^l [k,l]
} \nonumber \\
&&\times |r\rangle_{\cal E}
\sqrt{3^r [m^2 -k,r]
}
[m^2 -l -r, s-l-r],
\nonumber
\end{eqnarray}
where 
$|l\rangle_{\cal S}$ ($|r\rangle_{\cal E}$) are orthonormal states of a ``system'' \({\cal S}\) (``environment'' \({\cal E}\)) 
of \(k\) (\(m^2 -k\)) plaquettes.
When \(m^2 \to \infty\), the upper limit \(\min(s-l,m^2-k)\) is \(l\)-independent,
and
by Stirling approximation,
\(\ln [m^2 -t, s-t] 
\approx (m^2 -t) \ln (m^2-t) - (s-t) \ln (s-t)\)
plus functions of \(m^2\) and \(s\), where
$t=l+r$.
Linearizing the logarithm around the mean \(\overline{t}\) of \(t\) (in \(|\psi_{RVB}^d\rangle\)), and exponentiating
back, we obtain a product:
\([m^2 -t, s-t]
\approx q^l q^r\), with
\(q=(s-\overline{t})/(m^2-\overline{t}) \approx (-1+\sqrt{1+12d(1-d)})/(6(1-d))\).
The approximations are systematic, as \(\overline{t} \leq \sqrt{3}k/2\),
and
dispersion of \(t \sim \sqrt{\overline{t}}\).
%
%
%
So
the states of
\({\cal E}\) and \({\cal S}\) can be 
well  approximated as product states, which further
implies that 
\(|\Phi_{RVB}^d\rangle\),
for any 
\(d\), can be approximated  by product over
plaquettes, where each plaquette is 
in
the same pure
state -- a 
``mean-field picture'' is valid, for large \(m^2\). Consequently, the
entanglement of a system \({\cal S}\) of size \(k\), plus a boundary \(\mathbb{B}\) that
intersects \(\Delta\) plaquettes,
to the rest of the lattice, will scale as
\(hE_{pl}(d)+vE_{pl}(1-d)\),
where the boundary intersects \(h\) (\(v\)) plaquettes horizontally (vertically),
\(h+v=\Delta\), and
\(E_{pl}(d)= \log_2\left(1+3q^2\right)-\frac{3q^2}{1+3q^2}\log_2 q^2\).

For the most general initial state, 
\(|\Psi_0^{Hs}\rangle = \prod_{p=1}^{m^2}(|\psi_1\rangle|\psi_2\rangle|\psi_3\rangle|\psi_4\rangle)_p\),
denoting 
\(P_{\square,p}^{S=0}\) for the projector onto the space spanned by
\(\{|HH\rangle, |VV\rangle\}\) at the 
\(p^{\mbox{\footnotesize{th}}}\) plaquette, and \(\{|G_i\rangle\}\) for the ground states,
the cooled state is 
%
%
%
\(
\left(\sum_{i}|G_i\rangle \langle G_i| \right) |\Psi_0^{Hs}\rangle
%
= \left(\sum_{i}|G_i\rangle \langle G_i| \right)   \prod_{p=1}^{m^2} P_{\square,p}^{S=0} 
|\Psi_0^{Hs}\rangle
%
= \left(\sum_{i}|G_i\rangle \langle G_i| \right)
|\Psi_0^{H}\rangle
\),
so that the same area law holds.
Surprisingly therefore, the
area law depends on the \emph{path} of
the boundary (and not only on its length): it depends on \(h\), even for 
 fixed \(\Delta\).
However, the area law is linear in this 2D system, and supports our thesis for weakly frustrated systems.

\emph{Case 4: 
Shastry-Sutherland model.}
It is a 2D NN antiferromagnet (with Heisenberg couplings of strength \(J_1\)) on a square lattice \((i,j)\),
with additional Heisenberg interactions of strength \(J_2\) on the diagonals 
\((2i,2j) \leftrightarrow (2i+1,2j+1)\) and \((2i,2j+1) \leftrightarrow (2i-1,2j+2)\), with PBC.
%
%
The ground state of this model, 
for $J_1/J_2<1/2$, 
is a product of dimers along the \(J_2\) diagonals \cite{Shastry-Sutherland}. 
For $J_1/J_2<1/2$, by going to the Ising limit, the degeneracy becomes exponentially large, and \({\cal F} \approx 1/(1+ (1/2)(J_2/J_1))
< 1/2\), 
for large system size. 
Furthermore, the area law is linear,
and depends on the path of the boundary.


\emph{Case 5: 1D \(J_1\)-\(J_2\) Heisenberg model.}
Consider
a 1D system with PBC and \(2m\) sites,  having NN and NNN Heisenberg couplings of strengths \(J_1\) and \(J_2\) respectively.
%
%
For \(J_1 = 2J_2>0\)
(MG model \cite{Majumdar-ebong-Ghosh}),
%
the GSs are
\( |G_{MG}^{\pm}\rangle = \prod_{i =1}^{m} \left(|0\rangle_{2i} |1\rangle_{2i \pm 1} -
|1\rangle_{2i} |0\rangle_{2i \pm 1}\right)/\sqrt{2}\).
At the MG point, \({\cal F} = 1/2\). Note however that if we replace \(\vec{\sigma}_i\) by a 3D classical vector,
the frustration degree is zero.
%
%
Using the initial state \(\prod_{i=1}^{m-1} |0\rangle_{2i-1} |1\rangle_{2i}
|\phi\rangle_{2m-1} |\phi\rangle_{2m}\), where \(|\phi\rangle\) is an
arbitrary (qubit) state,
we obtain a \emph{lower bound}: \(E_{k:rest} \geq 2\) or \(1\), for even or odd \(k\).
Any cooled state 
is 
of the form \(|\Phi_{MG}\rangle = a|G_{MG}^+\rangle + b|G_{MG}^-\rangle\), whence,
after writing in Schmidt decompostion, an  upper bound reads:
 \(E_{k:rest} \leq \log_2 5\) or \(\log_2 3\), for even or odd \(k\).
%
Numerical simulations at the MG point, show that e.g. for \(2m =8\), \(E_{4:rest} \approx 2.314\) (cf. \cite{MGnumerical}), and
BE after cooling converges with \(k\).
Therefore, the area law is a constant for this 1D system, in support of our thesis.
Note that the above method of finding bounds can potentially be used in other models whose GS space is made up of dimers.

\emph{Case 6: Ising spin chain with NN interactions.}
The Hamiltonian, of \(2m\) spins (with PBC),
is \(H_{NN}^{Ising} =  \sum _{\langle ij \rangle} J_{ij} \sigma_i^z  \sigma_j^z\), with
all \(|J_{ij}| = J\), and all except one are negative.
\({\cal F} = 1/(2m-1)\).
The initial state 
$\ket{\Psi_0^{NN}}= \prod_i ((|0\rangle + |1\rangle)/\sqrt{2})_i$
(cf. \cite{somporko_thhakteo_parey}),
gives
the cooled state 
\(
|\Phi_{NN}^{Ising}\rangle = \frac{1}{2\sqrt{m}} \sum_{k=0}^{m}
[(|0^{\otimes (2m-k)} 1^{\otimes k}\rangle + |1^{\otimes (2m-k)} 0^{\otimes k}\rangle) +
(|1^{\otimes (k+1)} 0^{\otimes (2m -k -1)}\rangle + |0^{\otimes (k +1)} 1^{\otimes (2m -k -1)}\rangle)]\).
For any  \(k\), \(E_{k:2m-k}\) decreases with \(2m\),
contrary to previous cases.
Moreover,
\(E_{k:2m-k}\), for  a fixed \(2m\) 
 is a constant with \(k\) --
in support of our thesis.

\begin{figure}[t]
\includegraphics[width=7cm,height=3cm,clip=true]{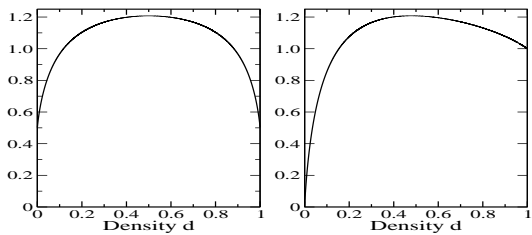}
\caption{Constructive and destructive interference. (Left)
$E/\overline{E}$ for a square \(\mathbb{B}\), so that $v=h$. (Right)
$E/\overline{E}$ for a horizontal (vertical) boundary, so that $v=0$ ($h=0$).
The vertical axes are of \(E/\overline{E}\).}
\label{entropysquare}
\end{figure}

\emph{Constructive and destructive interferences.}
The above studies allows us to identify an interesting interplay between frustration and interference of entanglement.
Beginning with \emph{Case 3}, for a fixed density \(d\),
the lattice is in the state $|\psi_{RVB}^d\rangle$, which is an equal superposition over
all states that have $s$ vertical plaquettes and
$m^2-s$ horizontal ones.
If we choose one of the states in this superposition,
calculate the entropy of the inner area (\({\cal S} \cup \mathbb{B}\)) in the chosen state, and then
average this entropy over all states in the superposition, the average entropy is
\(
\overline{E}(d)~=~2hd+2v(1-d)~.
\)
For a square \(\mathbb{B}\), the interference is destructive
(i.e. \(E/\overline{E} < 1\)) for low and high
densities, but otherwise constructive, and 
for a purely horizontal (or vertical) \(\mathbb{B}\), 
interference 
can wipe out the entanglement completely, at low densities (Fig. \ref{entropysquare}).
For a square \(\mathbb{B}\), \(E/\overline{E}\) is maximal for \(d=1/2\). Correspondingly, in the AF LR Ising gas,
the highest BE scaling is for \(\lambda =0\), when there are an equal number \(|0\rangle\)s and \(|1\rangle\)s in
the components of the cooled state,
and \({\cal F}\) is also maximal exactly at \(\lambda =0\).
We 
predict that a parent Hamiltonian that describes the ground states in \emph{Case 3} for fixed density, will be
maximally frustrated at \(d=1/2\).

For a similar study in the AF LR Ising gas (\emph{Case 1}), we normalize the entanglement scaling in the frustrated system, i.e. \(J>0\),
with the one in the nonfrustrated one (\(J<0\)). Since the latter is unity, the normalized entanglement, to leading 
order, is 
 \(\frac{1}{2}\log_2 k\), for large blocks of size \(k\), large system size \(2m\) (\(\gg k\)), and
for all \(\lambda\).
Thus we obtain constructive interference of BE  for all \(\lambda\), with respect to the nonfrustrated situation.
For the MG model, let us once again choose any one of the dimer ground states, find the entanglement in a bipartite split, and
average over the two ground states. The result is unity, for any split, so that we again have 
constructive
interference of entanglement.
For the system described in \emph{Case 6}, we have marginally constructive interference, compared to the nonfrustrated case (all \(J_{ij} <0\)).

In \emph{Case 2}, we have a rather remarkable example of destructive 
interference, because when \(2m\gg k\), \(E_{k:rest} \sim k\) for most elements in the superposition forming the cooled state, 
while the latter has only logarithmic scaling at most.
Superposition can therefore give rise to qualitative changes in scaling.


\emph{Summary.}
Firstly, our studies show that the
dimension of the GS manifold does not provide a ``good''  characterization of frustration.
This can be seen by comparing the $J_1-J_2$ model at the MG point
 with the 1D Ising model with a single frustrated bond.
Secondly, we found that trying to confer an ``area law'' on a frustrated system can have surprising consequences,
such as logarithmic divergence of BE in an effectively infinite dimensional system, and dependence of 
BE on the shape (and not only the Euclidean area) of the boundary.
Interestingly, the seeming independence of area law on dimension, in the long-range Ising model, gives rise to the possibility 
of applying density matrix renormalization group techniques to complex systems with long-range interactions (cf. \cite{Artur}).
Our results indicate that while weakly frustrated systems follow the usual area law known in the literature, strongly 
frustrated systems will each have their own area law. 
Finally, we have introduced 
a cooling procedure to study BE in degenerate ground state manifolds, a frustration degree, and a method to quantify 
constructive 
and destructive interference of entanglement.

We acknowledge comments of H.-U. Everts, R. Fazio,  T. Vekua, and support of
research project of Polish Government scientific funds (2005-08),
ESF QUDEDIS, 
Marie Curie ATK project COCOS (MTKD-CT-2004-517186),
Spanish MEC (FIS-2005-04627/01368, Consolider Project QOIT,
\& Ram{\'o}n y Cajal), \& EU IP SCALA.









\begin{thebibliography}{10}
\bibitem{footnote} If the scaling is not unique
after the maximization, we maximize  entanglement in the partition \((N/2) -1: (N/2)+1\), etc.,
until a unique scaling is obtained.
\end{thebibliography}
\end{document}